# Applications and Implications of Large Language Models in Qualitative Analysis: A New Frontier for Empirical Software Engineering

Matheus de Morais Leça
University of Calgary
Calgary, AB, Canada
matheus.demoraisleca@ucalgary.ca

Lucas Valença
University of Calgary
Calgary, AB, Canada
lucas.rodriguesvalen@ucalgary.ca

Reydne Santos
Universidade Federal de Pernambuco
Recife, PE, Brazil
rbs8@cin.ufpe.br

Ronnie de Souza Santos
University of Calgary
Calgary, AB, Canada
ronnie.desouzasantos@ucalgary.ca

***Abstract*—*Context*. The use of large language models for qualitative analysis is gaining attention in various fields, including software engineering, where qualitative methods are essential to understanding human and social factors. *Goal*. This study aimed to investigate how LLMs are currently used in qualitative analysis and how they can be used in software engineering research, focusing on identifying the benefits, limitations, and practices associated with their application. *Method*. We conducted a systematic mapping study and analyzed 20 relevant studies to explore reports of using LLM for qualitative analysis reported in the literature. *Findings*. Our findings indicate that LLMs are primarily used for tasks such as coding, thematic analysis, and data categorization, with benefits including increased efficiency and support for new researchers. However, limitations such as output variability, challenges capturing nuanced perspectives, and ethical concerns regarding privacy and transparency were also evident. *Discussions*. The study highlights the need for structured strategies and guidelines to optimize LLM use in qualitative research within software engineering. Such strategies could enhance the effectiveness of LLMs while addressing ethical considerations. *Conclusion*. While LLMs show promise for supporting qualitative analysis, human expertise remains essential for data interpretation, and continued exploration of best practices will be crucial for their effective integration into empirical software engineering research.**

## I. Introduction

Qualitative research, explores complex, context-based information to understand underlying meanings within a phenomenon [1]–[3]. This approach is not aimed at deeply exploring specific aspects of a research problem. Through qualitative methods, researchers aim to understand and represent the rich details that go beyond what can be quantified [3], [4].

Common methods in qualitative research allow for flexibility and close engagement with the subject matter. Grounded theory, for instance, involves gathering detailed data to inductively build theories rooted in participants' narratives [5], [6]. Ethnography enables researchers to immerse themselves within a community or organization, gaining insights into the social norms and cultural values that shape behavior [7], [8]. Action research emphasizes collaboration, encouraging researchers and participants to work together toward practical change [9], [10].

Qualitative analysis transforms these diverse forms of data into meaningful information through interpretation and categorization [11]. Researchers use the data by coding it into themes or patterns that reveal knowledge [2], [11], [12]. This process, often iterative, involves revisiting and refining categories, building up layers of meaning from the data itself, aiming to uncover a deeper understanding of experiences [12].

In software engineering, qualitative research plays an important role in exploring the human aspects of software development, such as team dynamics, communication styles, and organizational culture [2], [13]. As software projects depend on more than just technical skills—that is, they are shaped by the interactions and relationships within development teams—qualitative methods allow researchers to understand how these social factors influence software outcomes [8], [14].

The qualitative analysis process can be demanding due to the vast amount of unstructured data researchers must analyze and interpret. Manual strategies, such as hand-coding and organizing excerpts, allow for a thorough understanding but can be time-consuming and susceptible to errors [12], [15]. To support this, various tools offer automated coding, data organization, and visualization features, helping researchers manage data more efficiently and focus on interpreting their findings [16]–[19]. More recently, Large Language Models (LLMs) started to be incorporated in qualitative analysis in many fields, showing promising results [20]–[22], including in software engineering, but also raising challenges such as ethical and privacy concerns, especially with sensitive data being collected from software companies about their technologies that often are not released yet [20], [23].

Considering the growing use of large language models across various tasks and their potential impact on qualitative research methods, including in software engineering, our study focuses on investigating how researchers are using LLMs to support qualitative analysis. We aim to understand the current applications of these tools, identifying both their benefits and the challenges they present. By exploring the landscape of LLM usage in qualitative research, we expect to generate initial recommendations that could inform practical strategies for researchers, supporting the integration of these tools into research workflows. In this process, we seek to answer the following research question: ***How are researchers utilizing large language models to support qualitative analysis, and what are the reported benefits, challenges, and limitations of these tools in practice?***

The contributions of our research are threefold. First, we outline the current applications of LLMs in qualitative analysis, with a focus on methods such as coding, thematic analysis, and grounded theory. Second, we identify the specific LLM tools and techniques, such as prompt engineering, that researchers use to support qualitative data analysis, as well as the types of data on which LLMs are commonly applied, including interview transcripts and academic documents. Third, we address the challenges and ethical concerns related to LLM usage in qualitative research, particularly around privacy, consistency, and transparency, and suggest strategies for responsible integration of these tools into qualitative workflows.

## II. BACKGROUND

Large language models are advanced AI systems designed to process and generate text in ways that resemble human language use. Built on transformer architectures, they feature layers of neural networks that enable understanding of syntax, semantics, and context. These models learn by analyzing vast amounts of text data, often drawn from the Internet or curated sources, allowing them to recognize linguistic patterns, grasp subtle language structures, and generate contextually appropriate responses. Through this training, LLMs can perform tasks such as summarizing text, translating languages, answering questions, and even generating creative content or code. Their adaptability and extensive language capabilities make them valuable tools across a range of applications [24]–[26].

LLMs are now being used widely across research fields due to their potential to streamline text-heavy tasks [24], [27]. This usage naturally extends into qualitative research, where interpreting language and identifying themes are central [28]. LLMs can support coding and thematic analysis by recognizing patterns and organizing data, making them promising for tasks traditionally managed by human researchers [28]. Researchers can use LLMs to analyze large text datasets, potentially saving time and resources [29], [30]. However, while LLMs may enhance qualitative research, issues of reliability, consistency, and ethical concerns around privacy and data transparency remain important considerations for their effective integration [23], [31], [32]. As LLMs continue to advance, researchers will need strategies for quality control and reproducibility, as well as for identifying biases in LLM-generated data and addressing ethical concerns, to ensure that LLM-supported empirical studies uphold rigor and reliability.

## III. METHOD

In this research, we conducted a mapping study to explore and analyze published papers that used LLMs to support qualitative analysis. We maintained our focus on studies conducted in the context of software engineering. However, we extended our search for studies in other fields. To conduct this mapping study, we followed established guidelines and protocols for systematic reviews in software engineering [33], as detailed below.

**Specific Research Questions**. We developed three specific research questions that helped direct our data collection, analysis, and synthesis of evidence to offer a comprehensive overview of research available that discusses the use of LLMs for qualitative analysis. The questions are as follows: *RQ1*. How are LLMs used to support qualitative analysis? *RQ2*. What are the benefits of using LLMs for qualitative analysis? *RQ3*. What are the limitations associated with the use of LLMs in qualitative analysis?

**Search Strategy**. We started with a manual search that considered the leading conferences and journals on empirical methods in software engineering, which includes Evaluation and Assessment in Software Engineering (EASE), Empirical Software Engineering and Measurement (ESEM), Empirical Software Engineering Journal (EMSE), Journal of Systems and Software (JSS), International Conference on Software Engineering (ICSE) and Information and Software Technology (IST). In addition, we also manually searched for papers in leading journals about information technology and methodology, which included the Journal of Information Technology (JIT), Digital Society (DS), and the International Journal of Qualitative Methods (IJQM). We then created and applied a search string [1], which was used to identify papers in three of the most relevant digital libraries in software engineering: ACM, IEEE, and Scopus. As a result, we obtained 2,574 articles that could potentially be used to answer our research questions.

**Selection Process**. Following the guidelines from [33], to refine the 2,574 papers we retrieved from our manual and automatic search, we manually reviewed the lists of papers and assessed the title and abstract of each one. Based on that data alone, we applied five exclusion criteria and one inclusion criterion. According to the exclusion criteria, we filtered out papers that: 1) EXC1 - Do not discuss prompt engineering and LLMs in support of qualitative methodologies or analyses; 2) EXC2 - Are not full papers (e.g., master's dissertations,

[1](“Large Language Models” OR “LLMs” OR “GPT” OR “prompt engineering” OR “prompting”) AND (“qualitative analysis” OR “qualitative research” OR “qualitative methodology”)

doctoral theses, course completion monographs, short papers). As a rule of thumb, full papers must be at least 5 pages long; 3) EXC3 - Cannot be downloaded through university logins; 4) EXC4 - Are not written in English; 5) EXC5 - Are duplicate studies. After a closer examination of the remaining papers, we selected our final dataset by including (INC1) papers that focus on the use of LLMs in qualitative research methodologies (i.e., thematic analysis, literature review, interviews, content analysis, and case studies) or report their experiences in applying it. In this process, three independent researchers applied the exclusion and inclusion criteria. Any uncertainties were resolved through discussion between the three of them, with the option to involve a fourth researcher if necessary. This approach ensured that the final selection included papers that specifically addressed instances of generative AI in qualitative analysis. Finally, 20 papers were included. The article IDs and titles, as well as the coding process, are shown in Table I.

**Extraction, Analysis, and Synthesis**. After the inclusion and exclusion processes, two researchers reviewed the full text of the 20 included papers. The information extracted included the type of study, type of qualitative analysis supported by LLMs/AI, type of data analyzed using LLMs, benefits of using LLMs in qualitative analysis, limitations, reported threats to validity, tools used to conduct the qualitative analysis, and techniques applied.

Furthermore, we applied the thematic analysis method [34] to synthesize the collected data. First, two researchers immersed themselves in the data for familiarity and to identify segments relevant to the research. Second, we identified key concepts and created a list of codes. These codes were then translated into themes. During this step, similar codes were grouped together. Additionally, we evaluated the results of the synthesis to assess reliability. The results of this process are presented in the next section.

TABLE I: Analyzed Studies

| Id | Title |
|---|---|
| A001 | Applying Large Language Models to Interpret Qualitative Interviews in Healthcare (Wosny & Hastings, 2024) |
| A002 | Assessing ChatGPT's ability to emulate human reviewers in scientific research: A descriptive and qualitative approach (Suleiman et al., 2024) |
| A003 | The Role of Generative AI in Qualitative Research: GPT-4's Contributions to a Grounded Theory Analysis (Sinha et al., 2024) |
| A004 | Are Prompts All You Need?: Chatting with ChatGPT on Disinformation Policy Understanding (Chen et al., 2024) |
| A005 | Role Play: Conversational Roles as a Framework for Reflexive Practice in AI-Assisted Qualitative Research (Thominet et al., 2024) |
| A006 | Performing an Inductive Thematic Analysis of Semi-Structured Interviews With a Large Language Model: An Exploration and Provocation on the Limits of the Approach (Paoli, 2024) |
| A007 | CollabCoder: A Lower-barrier, Rigorous Workflow for Inductive Collaborative Qualitative Analysis with Large Language Models (Gao et al., 2024) |
| A008 | The use of Generative AI in qualitative analysis: Inductive thematic analysis with ChatGPT (Perkins & Roe, 2024) |
| A009 | Reflections on inductive thematic saturation as a potential metric for measuring the validity of an inductive thematic analysis with LLMs (Paoli & Mathis, 2024) |
| A010 | Enhancing the Analysis of Interdisciplinary Learning Quality with GPT Models: Fine-Tuning and Knowledge-Empowered Approaches (Zhong et al., 2024) |
| A011 | Prompts, Pearls, Imperfections: Comparing ChatGPT and a Human Researcher in Qualitative Data Analysis (Wachinger et al., 2024) |
| A012 | An Examination of the Use of Large Language Models to Aid Analysis of Textual Data (Tai et al., 2024) |
| A013 | Comparing GPT-4 and Human Researchers in Health Care Data Analysis: Qualitative Description Study (Li et al., 2024) |
| A014 | Artificial Intelligence and content analysis: the large language models (LLMs) and the automatized categorization (Carius & Teixeira, 2024) |
| A015 | Prompting Large Language Models for Topic Modeling (Wang et al., 2023) |
| A016 | Exploring the Use of Artificial Intelligence for Qualitative Data Analysis: The Case of ChatGPT (Morgan, 2023) |
| A017 | Me and the Machines: Possibilities and Pitfalls of Using Artificial Intelligence for Qualitative Data Analysis (Chubb, 2023) |
| A018 | Exploring the Use of AI in Qualitative Analysis: A Comparative Study of Guaranteed Income Data (Hamilton et al., 2023) |
| A019 | The Promise and Challenges of Using LLMs to Accelerate the Screening Process of Systematic Reviews (Huotala et al., 2024) |
| A020 | ChatGPT application in Systematic Literature Reviews in Software Engineering: an evaluation of its accuracy to support the selection activity (Felizardo et al., 2024) |

## IV. FINDINGS

Our review and analysis of the final 20 articles revealed an increasing interest of researchers in using LLMs mainly for coding tasks and thematic analysis. Other qualitative data analysis methods included grounded theory, article screening processes, content analysis, topic modeling, vignette analysis-creation, and critical review, as seen in Table II. Most of these procedures were done with ChatGPT, GPT-3.5, and GPT-4 by applying prompt engineering techniques. The relation between articles, LLM tools, and applied techniques can be seen in Tables III and IV.

**Usage**. Considering the context of qualitative analysis, coding strategies, and thematic analysis were the most widely supported methods by LLMs. Researchers often find LLMs useful for identifying themes and patterns efficiently, reducing the time, costs, and resources these methods usually require. LLMs have primarily been applied to analyze interviews and participant data, including focus groups, classroom transcripts, and students' posts, as well as articles, documents, and other textual data, such as university programs and policies on AI tools in academic research. The relationship between articles and types of data being analyzed by LLMs is shown in Table V.

LLMs have been reported to be valuable tools for coding tasks in qualitative research, offering fresh perspectives each time they are used. For example, each session with an LLM can act as a new code rater, providing varied insights that enrich the coding process and allow researchers to approach the data from multiple angles. This repeated coding ability makes LLMs especially useful for identifying data segments that human coders might miss, prompting further discussion on the data. Additionally, LLMs were demonstrated to assist in analyst triangulation, where different perspectives are compared to strengthen the reliability of findings. This capability not only supports experienced researchers in validating their interpretations but also serves as a learning tool for beginners, helping them build confidence and skills in coding and analysis.

TABLE II: Qualitative Data Analysis Methods

| Qualitative Data Analysis Method | Article ID |
|---|---|
| Coding (deductive, inductive, line-by-line) | A003, A004, A010, A012, A013, A016, A018 |
| Thematic analysis | A001, A005, A006, A007, A008, A009, A011 |
| Grounded theory | A003, A011 |
| Study screening | A019, A020 |
| Content Analysis | A014 |
| Topic modeling | A015 |
| Vignette analysis & creation | A017 |
| Critical review | A002 |

TABLE III: LLM Tools

| LLM Tool | Article ID |
|---|---|
| ChatGPT | A002, A004, A008, A010, A011, A012, A015, A016, A017, A018, A020 |
| GPT-3.5 | A001, A006, A007, A009, A019 |
| GPT-4 | A003, A007, A013, A014, A019 |
| Claude | A005 |
| LLaMA-13B | A015 |
| Mixtral 7x8b | A001 |

TABLE IV: LLM Techniques

| Applied Technique | Description | Article ID |
|---|---|---|
| Prompt Engineering | Prompt design techniques and queries to approach the best result. | A001, A002, A003, A004, A006, A007, A008, A009, A010, A011, A012, A013, A014, A015, A016, A017, A018, A019, A020 |
| Comparison and Evaluation | Result comparison with human reviews, metrics to measure results, performance thematic saturation. | A002, A007, A009 |
| Fine-Tuning and Parameter Adjustments | Fine-tuning and parameter adjustments such as temperature parameter, enhancing precision of answers. | A006, A008, A010 |
| Zero, One, or Few-Shot Learning | Zero-shot learning, One-shot, Few-shot learning, Few-shot with Chain-of-Thought prompting. | A004, A019 |

Our findings demonstrate that the usage of LLMs in qualitative analysis appears broad, extending across various scenarios and types of data. Researchers have been using LLMs to work with both structured and unstructured data in contexts ranging from educational settings to policy analysis, and for different data analysis tasks, such as initial data coding, thematic identification, and document screening processes.

**Benefits**. Our findings demonstrate several benefits of using LLMs in qualitative research, which are summarized in Table VI. *Support on identifying themes and patterns* is one prominent advantage, where LLMs assist in recognizing themes and patterns that may not be immediately apparent to human researchers. They facilitate comprehension of descriptive themes and enable a deeper understanding of complex qualitative data, helping researchers to capture patterns that might otherwise go unnoticed.

Another significant benefit is *efficiency and reduction of efforts*, as LLMs ease the cognitive demands on researchers and speed up traditionally time-consuming processes. They reduce the time, energy, and resources usually required in exhaustive analysis tasks, allowing researchers to allocate more focus on interpretation. *Support on analysis and coding* is also evident, as LLMs facilitate the coding process and assist in creating codebooks, generating foundational analytical content that supports further research.

Our findings highlight the *autonomy and flexibility* that LLMs offer researchers, particularly benefiting beginners by providing guidance and allowing for adjustments during the analysis process. However, this also requires researchers to critically assess the validity of the results. Future studies should further investigate whether novice researchers can reliably evaluate the consistency of LLM-generated outputs. Additionally, our findings suggest that LLMs enhance *human-to-human interaction* by facilitating collaboration during coding and analysis. These tools can assist by suggesting categorizations, summarizing discussions, and providing alternative perspectives, ultimately supporting the construction of a well-rounded discussion. Lastly, LLMs offer *support for triangulation*, where their non-deterministic nature complements human judgment in validating findings from multiple perspectives, thereby strengthening research reliability.

**Limitations.** Our findings reveal several limitations of using LLMs in qualitative research, as illustrated in Table VII. *Consistency, precision, and accuracy* are notable challenges, as LLMs often produce variable results due to their stochastic nature. The same prompt may yield different responses each time, and the models can be highly sensitive to prompt wording and input data. Additionally, LLMs are prone to generating ”hallucinations”, where they produce incorrect or entirely fabricated information. This lack of reliability makes it difficult for researchers to depend on LLM outputs for consistent analysis.

Another limitation lies in *high-level comprehension and critical analysis*, where LLMs struggle to capture nuanced meanings and subtle contexts and engage deeply with complex theoretical frameworks. While they can perform basic coding and thematic analysis, they lack reflexivity and cannot fully engage with intricate theoretical approaches that require interpretive depth. These limitations mean that human oversight remains essential for tasks requiring a deep, context-aware understanding.

*Ethics, privacy, and transparency* pose significant concerns in the use of LLMs for qualitative research. While these tools are often associated with efficiency gains, researchers must recognize that increased productivity cannot come at the expense of ethical considerations. LLMs operate without an internal ethical framework and do not obtain consent when processing provided data. Sharing sensitive participant information with LLMs raises serious privacy risks, particularly in regulated fields such as healthcare. The black-box nature of LLM decision-making and the lack of regulatory oversight introduce additional ethical challenges. There is no clear accountability for how these models process and interpret sensitive information, nor for the composition of their training data. Since LLMs are developed by private companies, they may inherently reflect and reinforce hegemonic knowledge. While some researchers view these tools as a means to reduce human biases, the underlying data composition may actually privilege dominant perspectives, potentially standardizing scientific inquiry.

LLMs face additional *technical limitations* as well, with token size constraints that restrict the model's capacity to process large datasets or lengthy interviews in one go. Their computational demands can also be resource-intensive, especially for extensive data, which may limit accessibility for researchers without substantial technical resources. Lastly, the rapid development of AI technology creates a risk of *technological and methodological dependency*. Researchers may become overly reliant on LLMs, while fast-paced advancements can render certain methodologies obsolete. LLMs also struggle to replicate human behaviors and interpretive nuances, which could reduce the richness of qualitative analysis if relied on exclusively.

## V. Discussions

As we compare our findings to the ones presented in the literature, we observe that the evidence we identified on the application of large language models in qualitative analysis aligns with some patterns seen in previous studies, such as the use of LLMs for coding and thematic analysis in various fields [28], [29]. Our findings also demonstrate that while LLMs enhance efficiency in processing unstructured data, they cannot handle the full depth of analysis independently. Literature shows that prompt engineering and carefully structured input can enhance the performance of LLMs [35], [36], a finding consistent with our evidence where contextually optimized commands were reported to improve LLMs' ability to generate relevant outputs. Additionally, like previous studies, we observed that ethical and privacy concerns remain crucial, especially when handling sensitive data that could impact data transparency and trustworthiness [23], [32].

However, we found certain differences from the recent discussions in the literature. For instance, while LLMs are often seen as tools that assist with well-defined tasks such as text summarization and sentiment analysis [24], [27], we found that they are increasingly being explored for more interpretive tasks, such as generating coding schemes and facilitating researcher collaboration. This flexibility to adjust LLM prompts according to qualitative needs suggests an evolution beyond prior goals, which mostly applied LLMs for straightforward NLP tasks without exploring complex qualitative research roles.

Our study contributes novel insights into how LLMs are integrated into qualitative research and how this can impact empirical software engineering. In particular, collected evidence indicates the importance of structured strategies tailored to qualitative research tasks in software engineering, where social dynamics and team interactions play an essential role. Additionally, the techniques we identified, e.g., prompt engineering, are promising practices that could be applied broadly to qualitative analysis. These findings demonstrate that as LLMs advance, new methodological frameworks will be needed to support their use in empirical software engineering research, encompassing prompt engineering (including different configurations of prompts, such as example-based prompting and chain-of-thought reasoning [35], [36]), quality control, and ethical considerations.

### A. Implications

Our findings suggest that, while the present state of LLM technology does not yet support fully autonomous analysis, these models can still serve as tools for assisting researchers with labor-intensive tasks, such as open coding, data categorization, and initial content analysis. These capabilities offer potential time savings and can support researchers in managing and interpreting large volumes of unstructured text data more efficiently.

As more advanced versions of LLMs are developed, their potential for handling complex qualitative analysis tasks will likely increase. By identifying and documenting the primary techniques currently used to facilitate qualitative analyses with LLMs, such as prompt engineering, this study provides a foundational understanding for researchers. Our study reveals that techniques, though applicable, come with limitations that should be acknowledged, and we provide some preliminary recommendations on addressing these issues.

One of the main implications of our study is that it highlights areas for future research on the limitations and capabilities of LLMs in qualitative analysis, encouraging refinement of techniques and exploration of new tools. As LLM technology advances, its applications and potential impacts on qualitative data analysis will continue to expand. Additionally, by categorizing commonly used techniques according to their suitability for specific types of analysis, such as thematic analysis or grounded theory, our study provides software engineering researchers with preliminary resources to make informed decisions about the use of LLMs in their studies.

### B. Recommendations

We compared the limitations of LLMs identified in our study with established guidelines for qualitative research and common threats to validity in software engineering research

TABLE V: Types of Data Analyzed Using LLMs

| Type of Data Analyzed Using LLMs | Description | Article IDs |
|---|---|---|
| Interviews and participant data | Interviews and focus groups transcripts, Classroom transcripts, Undergraduate students posts on Miro | A001, A003, A006, A009, A010, A011, A012, A013, A014, A016, A018 |
| Articles, documents and textual data | Articles, Book reviews, Master's university programs, Government documents, Policies about AI tools in academic research, Transcripts derived from arts-based journey maps | A002, A004, A005, A008, A017, A019, A020 |
| Researchers coding data | Data qualitatively coded by researchers during thematic analysis, content analysis or other procedures | A007 |
| Topic modeling general datasets | Datasets specifically designed for topic modeling | A015 |

TABLE VI: Benefits of Using LLMs in Qualitative Analysis

| Benefit of Using LLMs | Description | Article ID |
|---|---|---|
| Support on identifying themes and patterns. | Identification of themes and patterns not always apparent to humans. Facilitation of descriptive themes comprehension. Deepen comprehension of qualitative data. | A001, A003, A006, A011, A013, A015, A016, A017 |
| Efficiency and Reduction of Efforts. | Reduction of cognitive burden. Acceleration of exhaustive and long processes. Reduction of time, energy and resources usage. | A002, A003, A007, A008, A014, A016, A018, A019, A020 |
| Support on Analysis and Coding. | Facilitation of coding process and codebooks creation. Production of analytical content. | A001, A012, A014, A015 |
| Autonomy and Flexibility for Researchers. | Helpful for beginner researchers. Flexibility for researcher adjustements. | A004, A012, A020 |
| Contributions to Human-to-Human Interaction. | Enhancement of collaboration between participants during coding and analysis. | A005 |
| Support on Triangulation Process. | Non-deterministic approach is helpful for analysis, complementing human researcher triangulation. | A011 |

TABLE VII: Limitations of Using LLMs in Qualitative Analysis

| Limitation of Using LLMs | Description | Article IDs |
|---|---|---|
| Consistency, precision and accuracy | Non consistent results due to stochastic nature of LLMs. Non-deterministic results for the same prompt or question. Sensitive to prompt and input data. Hallucinations. | A001, A003, A004, A006, A008, A011, A012, A013, A014, A019, A20 |
| High-level comprehension and critical analysis | Difficulty achieving depth, subtle nuances and context. Lack of reflexivity. Difficulty engaging with complex theoretical approaches. | A001, A008, A011, A012, A014, A018 |
| Ethics, privacy and transparency | Lack of internal ethics, meaning and critical consciousness. Potential to return biased responses. Ethical and privacy concerns on sharing participant data. Potential harms because of wrong interpretations (e.g., medical context). Lack of transparency in internal work and decisions. Unregulated and unlegislated context. | A003, A005, A006, A008, A013, A017, A018 |
| Technical limitations | Token size limitations. Resource-intensive depending on data size. | A003, A006, A012, A013, A015 |
| Technological and methodological dependency | Over-reliance on researchers on AI. The rapid pace of tech evolution and the risk of making methodologies obsolete. Failures in emulating human reviews and behaviors. | A002, A007, A008 |

[14], [34], [37], [38]. Based on this comparison, we propose the following recommendations:

1) **Disclose LLM Tool and Version**: Authors should clearly specify the LLM tool and version used in their study. This disclosure supports result comparison and replication, as different versions may produce distinct outputs.
2) **Experiment with Prompt Strategies**: Researchers should test multiple prompt strategies to identify the most effective approach for their specific context [39]. For example, when analyzing interviews with software testers about fairness metrics, different types of prompts can enhance the analysis. *Instructional prompts* provide clear tasks by focusing on specific aspects, such as summarizing key concerns testers raise regarding fairness metrics. *Persona-based prompts* instruct the LLM to adopt a specific role, such as an ethics advisor, to assess fairness metrics and suggest improvements.

*Chain-of-thought prompts* encourage step-by-step reasoning, helping the model identify recurring themes, examine specific examples, and synthesize findings. Experimenting with these approaches can improve the quality, relevance, and depth of responses, maximizing the accuracy of the analysis. Additionally, to safeguard participant privacy and protect proprietary data from software companies, researchers should establish clear data protection protocols, including anonymizing data and ensuring that sensitive information is not directly input into LLMs.

3) **Establish Data Protection Protocols**: To safeguard participant privacy and protect proprietary data from software companies, researchers should implement clear data protection protocols, including anonymizing data and avoiding direct input of sensitive information into LLMs.
4) **Ensure Human Participation as a Fundamental Factor**: For tasks requiring critical analysis and interpretive depth, human oversight is essential. Researchers should not rely solely on LLMs for broad analysis, as these models lack the reflexivity needed for complex theoretical engagement and often reinforce uniform knowledge. In qualitative research, particularly in interviews, replacing human participants with LLMs risks limiting the depth of interpretation and diverse perspectives.
5) **Balance LLM Usage with Traditional Methods**: To prevent over-reliance on LLMs, researchers should integrate traditional qualitative methods to ensure that human interpretation remains a key part of the analysis process. This practice can be used as a validation step in the qualitative analyses.
6) **Follow Ethical Guidelines for AI Use**: Researchers should adhere to ethical guidelines tailored for AI applications, ensuring transparency in data handling, accountability, and respect for participant rights. This includes relying on open-source tools that do not operate solely as black boxes and implementing measures to protect participants' data privacy.
7) **Discuss Validity Threats Contextually**: Researchers should evaluate threats to validity in relation to their specific study goals and the LLM application, adapting threat discussion to align with the LLM unique configuration.

### C. Threats to Validity

Our research has limitations inherent to the method used, including the following. ***Internal Validity.*** Although our study selection followed predefined criteria and was conducted by three researchers, the analysis of titles and abstracts may have introduced bias, particularly in cases with unclear descriptions. The definition of codes and themes in thematic analysis depended on the researcher's judgment, potentially introducing personal biases. To address this, we adhered to established guidelines and conducted agreement discussions to reach a consensus when needed. ***External Validity.*** Despite employing two search strategies, the second (automatic search) posed a threat. The ACM search returned approximately 3,000 documents, yet only the top 1,000 could be exported, possibly limiting study diversity and scope. To mitigate this, we supplemented our search with additional libraries and a manual search in major software engineering conferences and journals. ***Conclusion Validity.*** Our study's conclusions are limited by the sample size of 20 articles, which may not robustly support our research questions, affecting the generalizability of findings on practices and recommendations for LLM use. Thus, we provide analytical procedures that enable researchers to derive context-specific insights rather than broad generalizations.

## VI. Conclusion

Our study investigated the use of LLMs in qualitative analysis, aiming to understand how these tools are currently applied and to identify practices, benefits, and limitations relevant to their integration. We focused on how findings from this investigation could be applied to software engineering research, given the reliance on qualitative methods within the field. Through a systematic mapping study, we reviewed the current literature to provide insights into the applications of LLMs in qualitative analysis, particularly regarding their potential to support researchers with tasks traditionally requiring human interpretation.

We identified that LLMs are applied primarily in tasks like coding, thematic analysis, and data categorization, providing efficiency by reducing the time, cognitive demands, and resources often required for these processes. Their benefits include support in identifying patterns, simplifying coding efforts, and offering guidance to researchers who are newer to qualitative analysis. However, we observed limitations, such as variability in outputs due to the stochastic nature of LLMs, difficulties in capturing nuanced understanding, and concerns around privacy and transparency. These findings suggest that while LLMs can assist in qualitative analysis, human expertise remains essential for in-depth interpretation.

In conclusion, our study highlights the need for structured strategies for incorporating LLMs into empirical software engineering research. Developing clear guidelines for using LLMs can help researchers integrate these tools effectively, making qualitative analysis more manageable while addressing ethical considerations. Establishing these strategies will be helpful for future studies, enabling LLMs to be used as supportive tools in software engineering and other areas reliant on qualitative data analysis.

## VII. Dataset Availability

The dataset with all the excluded and included papers, as well as the coded extracted data, is available here.

## References

[1] S. Sofaer, "Qualitative methods: what are they and why use them?" *Health services research*, vol. 34, no. 5 Pt 2, p. 1101, 1999.

[2] C. B. Seaman, "Qualitative methods," in *Guide to advanced empirical software engineering*. Springer, 2008, pp. 35–62.

[3] J. Corbin and A. Strauss, *Basics of qualitative research*. sage, 2015, vol. 14.
[4] M. Sandelowski, "Qualitative analysis: What it is and how to begin," *Research in nursing & health*, vol. 18, no. 4, pp. 371–375, 1995.
[5] K. Charmaz, "Grounded theory," *Qualitative psychology: A practical guide to research methods*, vol. 3, pp. 53–84, 2015.
[6] K.-J. Stol, P. Ralph, and B. Fitzgerald, "Grounded theory in software engineering research: a critical review and guidelines," in *Proceedings of the 38th International conference on software engineering*, 2016, pp. 120–131.
[7] J. Brewer, *Ethnography*. McGraw-Hill Education (UK), 2000.
[8] H. Sharp, Y. Dittrich, and C. R. De Souza, "The role of ethnographic studies in empirical software engineering," *IEEE Transactions on Software Engineering*, vol. 42, no. 8, pp. 786–804, 2016.
[9] D. E. Avison, F. Lau, M. D. Myers, and P. A. Nielsen, "Action research," *Communications of the ACM*, vol. 42, no. 1, pp. 94–97, 1999.
[10] P. S. M. Dos Santos and G. H. Travassos, "Action research can swing the balance in experimental software engineering," in *Advances in computers*. Elsevier, 2011, vol. 83, pp. 205–276.
[11] D. Ezzy, *Qualitative analysis*. Routledge, 2013.
[12] J. Ritchie, L. Spencer, W. O'Connor *et al.*, "Carrying out qualitative analysis," *Qualitative research practice: A guide for social science students and researchers*, vol. 2003, pp. 219–62, 2003.
[13] P. Lenberg, R. Feldt, L. Gren, L. G. Wallgren Tengberg, I. Tidefors, and D. Graziotin, "Qualitative software engineering research: Reflections and guidelines," *Journal of Software: Evolution and Process*, vol. 36, no. 6, p. e2607, 2024.
[14] C. B. Seaman, "Qualitative methods in empirical studies of software engineering," *IEEE Transactions on software engineering*, vol. 25, no. 4, pp. 557–572, 1999.
[15] G. R. Gibbs, "Qualitative analysis," *Qualitative Data Analysis*, p. 277, 2014.
[16] V. Talanquer, "Using qualitative analysis software to facilitate qualitative data analysis," in *Tools of chemistry education research*. ACS Publications, 2014, pp. 83–95.
[17] R. TESCH, "Computer software and qualitative analysis: A reassessment," *New Technology in Sociology: Practical Applications in Research and Work*, 2019.
[18] F. Freitas, J. Ribeiro, C. Brandão, L. P. Reis, F. N. de Souza, and A. P. Costa, "Learn for yourself: The self-learning tools for qualitative analysis software packages," *Digital Education Review*, no. 32, pp. 97–117, 2017.
[19] L. S. Gilbert, K. Jackson, and S. Di Gregorio, "Tools for analyzing qualitative data: The history and relevance of qualitative data analysis software," *Handbook of research on educational communications and technology*, pp. 221–236, 2014.
[20] J. Roberts, M. Baker, and J. Andrew, "Artificial intelligence and qualitative research: The promise and perils of large language model (llm)'assistance'," *Critical Perspectives on Accounting*, vol. 99, p. 102722, 2024.
[21] R. H. Tai, L. R. Bentley, X. Xia, J. M. Sitt, S. C. Fankhauser, A. M. Chicas-Mosier, and B. G. Monteith, "An examination of the use of large language models to aid analysis of textual data," *International Journal of Qualitative Methods*, vol. 23, p. 16094069241231168, 2024.
[22] M. Bano, D. Zowghi, and J. Whittle, "Ai and human reasoning: Qualitative research in the age of large language models," *The AI Ethics Journal*, vol. 3, no. 1, 2023.
[23] M. Bano, R. Hoda, D. Zowghi, and C. Treude, "Large language models for qualitative research in software engineering: exploring opportunities and challenges," *Automated Software Engineering*, vol. 31, no. 1, p. 8, 2024.
[24] I. Grossmann, M. Feinberg, D. C. Parker, N. A. Christakis, P. E. Tetlock, and W. A. Cunningham, "Ai and the transformation of social science research," *Science*, vol. 380, no. 6650, pp. 1108–1109, 2023.
[25] Y. Shen, L. Heacock, J. Elias, K. D. Hentel, B. Reig, G. Shih, and L. Moy, "Chatgpt and other large language models are double-edged swords," p. e230163, 2023.
[26] Y. Chang, X. Wang, J. Wang, Y. Wu, L. Yang, K. Zhu, H. Chen, X. Yi, C. Wang, Y. Wang *et al.*, "A survey on evaluation of large language models," *ACM Transactions on Intelligent Systems and Technology*, vol. 15, no. 3, pp. 1–45, 2024.
[27] H. Schroeder, M. A. L. Quéré, C. Randazzo, D. Mimno, and S. Schoenebeck, "Large language models in qualitative research: Can we do the data justice?" *arXiv preprint arXiv:2410.07362*, 2024.
[28] S. A. Gebreegziabher, Z. Zhang, X. Tang, Y. Meng, E. L. Glassman, and T. J.-J. Li, "Patat: Human-ai collaborative qualitative coding with explainable interactive rule synthesis," in *Proceedings of the 2023 CHI Conference on Human Factors in Computing Systems*, 2023, pp. 1–19.
[29] J. Gao, Y. Guo, G. Lim, T. Zhang, Z. Zhang, T. J.-J. Li, and S. T. Perrault, "Collabcoder: a lower-barrier, rigorous workflow for inductive collaborative qualitative analysis with large language models," in *Proceedings of the CHI Conference on Human Factors in Computing Systems*, 2024, pp. 1–29.
[30] L. Soldaini, R. Kinney, A. Bhagia, D. Schwenk, D. Atkinson, R. Authur, B. Bogin, K. Chandu, J. Dumas, Y. Elazar *et al.*, "Dolma: An open corpus of three trillion tokens for language model pretraining research," *arXiv preprint arXiv:2402.00159*, 2024.
[31] M. Hosseini and S. P. Horbach, "Fighting reviewer fatigue or amplifying bias? considerations and recommendations for use of chatgpt and other large language models in scholarly peer review," *Research integrity and peer review*, vol. 8, no. 1, p. 4, 2023.
[32] É. Ollion, R. Shen, A. Macanovic, and A. Chatelain, "The dangers of using proprietary llms for research," *Nature Machine Intelligence*, vol. 6, no. 1, pp. 4–5, 2024.
[33] B. A. Kitchenham, T. Dyba, and M. Jorgensen, "Evidence-based software engineering," in *Proceedings. 26th International Conference on Software Engineering*. IEEE, 2004, pp. 273–281.
[34] D. S. Cruzes and T. Dyba, "Recommended steps for thematic synthesis in software engineering," in *2011 international symposium on empirical software engineering and measurement*. IEEE, 2011, pp. 275–284.
[35] T. Heston and C. Khun, "Prompt engineering in medical education," *International Medical Education*, vol. 2, pp. 198–205, 8 2023.
[36] J. White, Q. Fu, S. Hays, M. Sandborn, C. Olea, H. Gilbert, A. Elnashar, J. Spencer-Smith, and D. C. Schmidt, "A prompt pattern catalog to enhance prompt engineering with chatgpt," *arXiv preprint arXiv:2302.11382*, 2023.
[37] P. Runeson and M. Höst, "Guidelines for conducting and reporting case study research in software engineering," *Empirical software engineering*, vol. 14, pp. 131–164, 2009.
[38] R. Verdecchia, E. Engström, P. Lago, P. Runeson, and Q. Song, "Threats to validity in software engineering research: A critical reflection," *Information and Software Technology*, vol. 164, p. 107329, 2023.
[39] Y. Chen, C. Wong, H. Yang, J. Aguenza, S. Bhujangari, B. Vu, X. Lei, A. Prasad, M. Fluss, E. Phuong *et al.*, "Assessing the impact of prompting, persona, and chain of thought methods on chatgpt's arithmetic capabilities," *arXiv preprint arXiv:2312.15006*, 2023.